\documentclass[conference]{IEEEtran}
\IEEEoverridecommandlockouts
\usepackage{cite}
\usepackage{amsmath,amssymb,amsfonts}
\usepackage{algorithm}
\usepackage{algpseudocode}
\usepackage{graphicx}
\usepackage{textcomp}
\usepackage{xcolor}
\def\BibTeX{{\rm B\kern-.05em{\sc i\kern-.025em b}\kern-.08em
    T\kern-.1667em\lower.7ex\hbox{E}\kern-.125emX}}
\begin{document}

\title{Privacy-Preserving Feature Valuation in Vertical Federated Learning Using Shapley-CMI and PSI Permutation\\
}

\author{\IEEEauthorblockN{Unai Laskurain}
\IEEEauthorblockA{\textit{Electronics and Computing Department} \\
\textit{Mondragon Unibertsitatea}\\
Arrasate-Mondragon, Spain \\
unai.laskurain@alumni.mondragon.edu}
\and
\IEEEauthorblockN{Aitor Aguirre-Ortuzar}
\IEEEauthorblockA{\textit{Electronics and Computing Department} \\
\textit{Mondragon Unibertsitatea}\\
Arrasate-Mondragon, Spain \\
aaguirre@mondragon.edu}
\and
\IEEEauthorblockN{Urko Zurutuza}
\IEEEauthorblockA{\textit{Electronics and Computing Department} \\
\textit{Mondragon Unibertsitatea}\\
Arrasate-Mondragon, Spain \\
uzurutuza@mondragon.edu}

}

\maketitle

\begin{abstract}
Federated Learning (FL) is an emerging machine learning paradigm that enables multiple parties to collaboratively train models without sharing raw data, ensuring data privacy. In Vertical FL (VFL), where each party holds different features for the same users, a key challenge is to evaluate the feature contribution of each party before any model is trained, particularly in the early stages when no model exists. To address this, the Shapley-CMI method was recently proposed as a model-free, information-theoretic approach to feature valuation using Conditional Mutual Information (CMI). However, its original formulation did not provide a practical implementation capable of computing the required permutations and intersections securely. This paper presents a novel privacy-preserving implementation of Shapley-CMI for VFL. Our system introduces a private set intersection (PSI) server that performs all necessary feature permutations and computes encrypted intersection sizes across discretized and encrypted ID groups, without the need for raw data exchange. Each party then uses these intersection results to compute Shapley-CMI values, computing the marginal utility of their features. Initial experiments confirm the correctness and privacy of the proposed system, demonstrating its viability for secure and efficient feature contribution estimation in VFL. This approach ensures data confidentiality, scales across multiple parties, and enables fair data valuation without requiring the sharing of raw data or training models.
\end{abstract}

\begin{IEEEkeywords}
Vertical Federated Learning, Data Valuation, Shapley Value, Conditional Mutual Information, Private Set Intersection, Feature Importance.
\end{IEEEkeywords}

\section{Introduction}
Federated Learning (FL) \cite{zhang_survey_2021} is an emerging collaborative machine learning paradigm that enables various data owners to train a shared model without sharing their raw data, thereby preserving data privacy. Depending on the distribution of local data, there are two main categories of FL scenarios: 1) Horizontal Federated Learning (HFL) and 2) Vertical Federated Learning (VFL). In HFL, participants share the same feature space but have different data samples (e.g., different users with the same data types) \cite{yang_horizontal_2020}, while in VFL, participants hold different feature spaces for the same set of entities (e.g., other institutions owning complementary information about the same users) \cite{liu_vertical_2024}. FL enables secure, decentralized learning across organizations or devices, using techniques such as encryption and secure aggregation to protect sensitive data.

Although most FL studies focus on HFL, VFL is an even more attractive implementation for many companies due to its business value, especially in settings where different organizations have complementary information about the same users. VFL is essential for enabling institutions that cannot share raw data, but need data owned by other institutions to train a collaborative model. In contrast, the data of all members remains private. For example, a bank and an e-commerce platform may want to collaborate to create a fraud detection model, where the bank would have access to features such as income, savings, and loan history, and the e-commerce platform could contribute features like spending patterns or purchase frequency. Thus, VFL becomes highly relevant in real-world applications, such as finance, healthcare, and public services, where different organizations hold valuable information for others, but the data must remain private and confidential.

Privacy is one of the pillars of VFL, as the model must be trained with raw data without directly sharing it. However, a primary challenge arises when starting to build the model, which is evaluating the contribution of each party's features to the final model without revealing sensitive data. Feature evaluation is essential for identifying which features are helpful for the model and which ones contain redundant or irrelevant data, enabling the determination of how much one party should compensate the other(s) based on the contribution their data makes to the model. As a result, designing secure methods to estimate feature importance becomes a critical task in making VFL practical and trustworthy.

Recent works, such as \cite{han_data_2021}, have attempted to address this problem by estimating the contribution of features based on Shapley values. Typically, Shapley values require a model to be calculated. Nevertheless, this method utilizes Conditional Mutual Information (CMI) to calculate the correlation between the features and the label, thereby obtaining an approximation of the Shapley value, referred to as Shapley-CMI. This approach involves creating permutations and calculating the size of the intersections of each possible combination of values, which is achieved using a Private Set Intersection (PSI) server, as described in \cite{christin_scaling_2014}. However, this method has some limitations, as it requires creating many permutations of features, where each one is created by adding features one by one, and calculating the intersection of all possible combinations of values. The problem with this approach is that the proposed PSI server itself does not create any permutation; it only computes the intersection of the identifiers (ID) that it receives, and the parties cannot create the permutations either, as they cannot see what features the others have. As a consequence, there is no one capable of creating these permutations, so parties do not know by themself which ID set they need to send at each moment, even though they are supposed to send ID sets and receive the intersection size with the ID sets of other parties as a result. Even if the PSI server were able to create the permutations, it could not ask the parties to provide the corresponding ID set in each iteration without knowing which data they have.

To overcome the limitations mentioned above, we propose a method that improves the logic of the system by making the parties discretize their values, grouping the IDs with the same discretized values, and sending all the ID groups to the PSI server, so that it can create the permutation based on the received ID groups. We also propose a method to reduce the number of iterations needed for each permutation, allowing the PSI server to compute only the intersections of the combinations of values the IDs have, rather than computing all possible combinations of values. Finally, we also propose a one-way encryption method for the IDs, ensuring that the PSI server is unaware of the identities and cannot decrypt them.

\section{Related work}
Several previous works have been dedicated to participant selection in federated learning, but most of them focus only on HFL. An example of this is a utility-aware participant selection mechanism based on data value estimation in HFL proposed in \cite{zhao_efficient_2021}. Their method utilizes Shapley values to estimate the contribution of each client; however, it does not apply to VFL settings where parties hold different features of the same data samples. Similarly, \cite{fan_improving_2022} introduces a reinforcement learning-based approach for client selection, but their work is also limited to horizontal scenarios.

Regarding the previous work done on data valuation in VFL, most of it relies on model-dependent metrics, such as feature importance or loss reduction. For instance, the work presented in \cite{fan_fair_2022} proposes a data valuation method by analysing the impact of each party's features on the model performance. However, their approach requires access to the trained model and relies on backpropagation to compute feature contributions. Some methods propose computing data values in VFL by measuring the contribution of each party's feature through gradient-based sensitivity analysis \cite{jiang_vf-ps_2022}. Other model-dependent methods, such as \cite{li_fedsdg-fs_2023}, leverage influence functions to estimate each party's contribution to the final model's performance. Still, it cannot be applied in scenarios where the model has not already been built. Similar to some of the HFL methods, approaches based on Shapley values, such as the ones proposed by \cite{song_profit_2019} and \cite{cui_survey_2024}, define data value according to each participant's marginal contribution to the model's performance. As a result, the dependence on model training makes them unsuitable in early-stage settings where the model is not yet available, as well as in privacy-sensitive scenarios where access to the whole model or repeated training is impractical. From a business perspective, understanding in advance how other parties' features can influence the model has a significant impact.

Even if most feature selection techniques depend on training a model, some can be performed in the preprocessing phase without building any model. The research in \cite{khan_vfl-rps_2025}  aims to conduct participant selection by measuring the Spearman correlation between the features. Yet, this method does not consider the contribution if another party has already provided the same feature, which is a characteristic we want to include for fair contribution estimation. Another technique for feature selection is the one proposed in \cite{li_privacy-preserving_2021}, which introduces a protocol for private feature selection using Gini impurity within a Secure Multiparty Computation framework. However, this approach does not account for feature correlations, such as when two or more features are similar or share the same data.

As mentioned above, Shapley values are often used to evaluate the contribution of different features during model training. However, \cite{han_data_2021} presents a method in which Shapley values are calculated using CMI rather than relying on model gradients or predictions. To calculate the CMI, the method proposes a private set intersection (PSI) server, similar to the ones applied in \cite{dong_fair_2013} and \cite{christin_scaling_2014}, where clients send their encrypted IDs and the server returns the cardinality of the intersection, rather than the IDs that make up the intersection. This method allows for estimating the importance of each party's feature in a VFL setting before the model is trained, but it has some limitations. To calculate the Shapley-CMI value, many permutations of the features are necessary. In each permutation, each feature is added sequentially, and the combinations of possible values are calculated. As a result, a method is required where the server is also able to generate all those permutations and calculate the intersections for each combination, without the parties knowing which and how many features the others have.

Concerning the privacy required in VFL settings, \cite{bonesana_flotta_2024}, \cite{cortes-mendoza_training_2024}, \cite{hou_balancing_2024} explore various security mechanisms such as secure aggregation, homomorphic encryption during model training, and privacy-preserving protocols for feature embedding. While these approaches enhance data protection during and after model training, they are not applicable in scenarios where data needs to be protected before any training occurs, such as in early-stage data valuation or participant selection based on raw feature contributions. Therefore, a privacy method is required to encrypt the IDs sent from clients to the PSI server, which does not require the ability to decrypt the encrypted data.

\section{Methodology}
In this section, we describe the methodology used in our system. First, we present the overall architecture, explaining the roles of the client parties and the PSI server, as well as the communication between them. Then, we explain how Shapley-CMI values are calculated using random permutations and PSI-based intersection counts. Finally, we describe the functioning of the clients and the PSI server, and how they interact to calculate the Shapley value.

\subsection{System architecture}\label{SA}
\begin{figure}[b]
\centerline{\includegraphics[width=\linewidth]{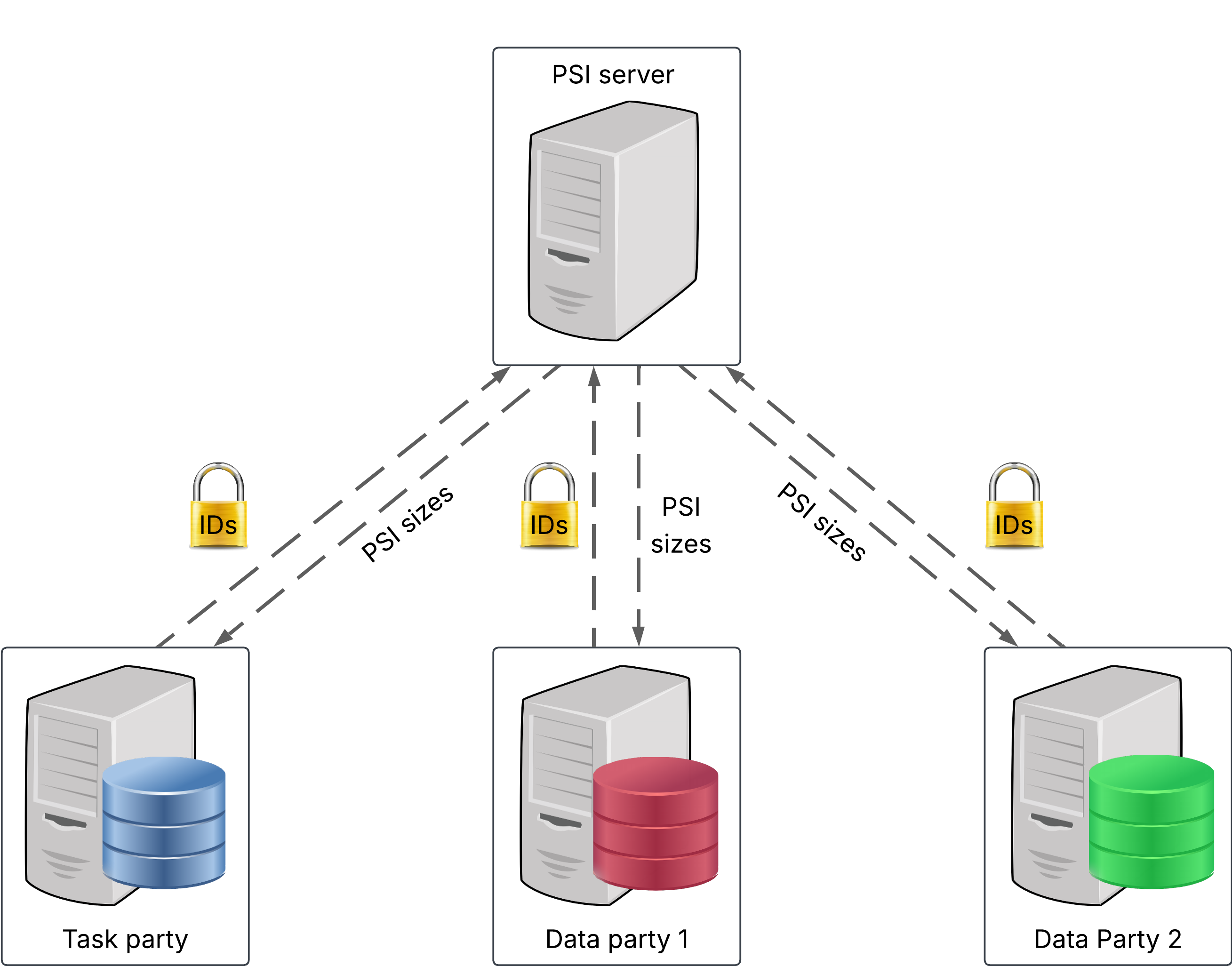}}
\caption{Architecture of the PSI server system.}
\label{fig:architecture}
\end{figure}

In this system, the architecture is composed of multiple client parties and a single PSI server. Each party holds a different subset of features for the same set of users, while the label is held by one of the parties. The party with the label will be referred to as the task party, while the other parties that contribute data will be referred to as data parties. The architecture of the system can be found in Fig.~\ref{fig:architecture}.

The communication process is organised in several steps. First, each party preprocesses its data locally by encrypting each ID sample with the HMAC-SHA256 method and discretizing feature values. This ensures that both the identity of the users and the raw values of the features are protected before any information is transmitted. Each of them ends up with lists of encrypted ID sets for each feature they own, which are sent to the central PSI server.

The PSI server has several responsibilities. The first one is to receive and combine all the ID lists of the parties and identify the users they all have in common. This identification step is necessary, as it ensures that all subsequent operations are performed only on shared user data, preventing any party from learning about users not common to all. After that, it creates random permutations of the order in which the features can be added to the training of the VFL model. For each one of those, it adds each feature one by one and calculates the size of PSI intersections of all possible group combinations. This process simulates the different marginal contributions that each feature might have when added in different orders, which is necessary to compute their Shapley-CMI values later. Finally, it collects all the intersection sizes for each feature and sends them back to the corresponding parties.

Using the intersection sizes sent by the server, each party calculates the CMI values locally and then averages them to obtain the Shapley-CMI value for each feature. This decentralised calculation step prevents the PSI server from accessing any numerical Shapley-CMI values derived from the data. This architecture ensures that no party ever learns about the raw data or features of others, and also guarantees that the PSI server operates only on encrypted and aggregated data, thereby preserving privacy at all times.

\subsection{Data valuation metric}
As mentioned above, the PSI server is responsible for computing the intersections of IDs, computing four different intersections for each possible combination of values. Thus, to understand what these four intersections are, where they come from, and how they are used to calculate the CMI, it is necessary to explain the basis of the Shapley-CMI metric. In this work, we estimate the contribution of each feature to the final model using the Shapley-CMI framework \cite{han_data_2021}, which combines game theory of Shapley and information theory of CMI. The value of Shapley of a feature is calculated as follows:
\begin{equation}
\phi_d = \sum_{D \subseteq \mathcal{D} \setminus \{d\}} \frac{|D|! \, (|\mathcal{D}| - |D| - 1)!}{|\mathcal{D}|!} \, I(X_d; Y_t \mid X_D X_t)
\label{eq:shapley-cmi}
\end{equation}

where \(\phi_d\) represents the Shapley value of feature \textit{d}, which measures its marginal contribution to the predictive power of the model. The summation runs over all features in \textit{D}, which is the set of features of the parties that have joined the game and yielded the game value, excluding the party \textit{d}. These parties belong to \(\mathcal{D}\), which are the features of all sets of parties in a game. The term \(I(X_d;\, Y_t \mid X_D X_t)\) denotes the CMI between feature \textit{d} and the target variable \(Y_t\), given the features of the other parties \(X_D\) and the features of task party \(X_t\).

Regarding the CMI, the formula is based on counting co-occurrences of discretized feature values and computing a conditional mutual information score. It is defined as follows:
\begin{equation}
\resizebox{\columnwidth}{!}{$
\begin{array}{c}
I(X_d;Y_t \mid X_D X_t) = \frac{1}{n} \sum_{x_d, x_D, x_t, y_t} N(x_d x_D x_t y_t) \\[0.7em]
\cdot \log \frac{
N(x_d x_D x_t y_t)
\sum_{\substack{x_d' \in \mathcal{X}_d,\, y_t' \in \mathcal{Y}_t}} N(x_d' x_D x_t y_t')}
{
\sum_{y_t' \in \mathcal{Y}_t} N(x_d x_D x_t y_t') 
\sum_{x_d' \in \mathcal{X}_d} N(x_d' x_D x_t y_t)
}
\end{array}
$}
\label{eq:cmi}
\end{equation}

where \(N(x_dx_Dx_ty_t)\) represents the empirical count of occurrences of the joint event \((x_d,x_D,x_t,y_t)\) in the dataset, where \(n\) represents the total number of samples. The numerator captures the joint probability of observing \(x_d,x_D,x_t\) and \(y_t\), scaled by the sum over all possible combinations of \(x'_d\) and \(y'_t\), reflecting the overall co-occurrence pattern. The denominator normalizes this by considering the marginal probabilities of \(x_d\) and \(y_t\) given \(x_D,x_t\), as well as considering \(y_t\) given \(x_d,x_D\) and \(x_t\), and \(x_d\) given \(x_D,x_t\) and \(y_t\). These four empirical counts are the four intersections the PSI server computes for each combination of values, and that is why they need to be sent back to the parties, so that they can use them to apply the formula \eqref{eq:cmi} and calculate the CMI of the feature.

To calculate the Shapley-CMI value using \eqref{eq:shapley-cmi}, permutations are necessary to fairly evaluate the contribution of each feature by considering all possible orders in which features can be added. This ensures that no feature is advantaged or disadvantaged by its position, leading to a balanced and accurate measure of its predictive power. However, as the number of permutations increases factorially with the number of features, it becomes computationally infeasible to evaluate all of them for large feature sets. To address this, this method uses a random subset of permutations to obtain an approximation of the actual Shapley-CMI value, making computation viable while still providing a reliable estimate of the contribution of each feature.

\subsection{Data preprocessing}
Before the feature contribution estimation process of the PSI server can begin, each party must preprocess its data to ensure privacy and the server's correct functioning. This preprocessing consists of two main steps: encrypting the IDs using a secure one-way hashing method and discretizing the feature values into ID groups. These steps are essential for enabling secure multiparty computation and avoiding any data leakage during communication or intersection operations.

First, each party encrypts the identifiers (IDs) of its data samples using the HMAC-SHA256 hashing function \cite{azeez_achieving_2018}. This technique combines the SHA256 cryptographic hash function with a secret key via HMAC to produce a unique and irreversible hash for each ID. The secret key would be previously shared securely by the task party with all the data parties, so that no one else has access to it. This sharing is assumed to be done through a secure channel such as public-key cryptography or a secure key exchange protocol. Since SHA256 is a one-way function, the original ID cannot be recovered from its hash. The encryption process is performed in four steps with this method. First, it encrypts the IDs by applying the secret key to encrypt the ID, and then it computes the hash (SHA256) of that encrypted result. Afterwards, it repeats the process by encrypting the obtained hash result with the secure key and computing the hash of that new encryption result. This double application of HMAC-SHA256 enhances resistance to brute-force attacks and dictionary attacks \cite{raza_survey_2012}, thereby reinforcing privacy even in adversarial settings. All this ensures that the identifiers remain private and anonymous throughout the protocol, even during communication with the PSI server and the computation of intersections.

Once encrypted, the encrypted IDs are grouped based on the discretized values of the corresponding features. Each feature is divided into a predefined number of intervals (or bins), where each bin represents a range of close values. For example, if a feature contains values between 1 and 20, it can be split into five bins such as [1–4], [5–8], [9–12], [13–16], and [17–20]. Then, the encrypted IDs of the samples are grouped into the bins their value belongs to, so that each of the bins contains a list of all the encrypted IDs whose value is between that range. The number of bins can be adjusted according to the required granularity or the sensitivity of the feature. As a result, for each feature they own, every party ends up with a collection of hashed ID sets, each corresponding to a specific interval of values.

The collection of ID sets is then sent to the PSI server via HTTPS protocol, where each feature is labeled with a predefined name (e.g., f1, f2, f3, etc.) and an identifier indicating which party sent it. The use of HTTPS guarantees transport-level encryption, preventing data leakage during data transmission. This allows the server to correctly track and manage data from different parties without knowing which features are being sent and who sent them.

\subsection{PSI intersection computation}
The PSI server receives all encrypted ID groups of the features from all parties and, once it has received all of them, merges them into a single dictionary that contains the list of ID groups for all features. This unified structure enables the server to systematically organize the input data from different parties and prepare it for processing. VFL assumes that all parties share the same set of ID samples; however, parties can also contain additional IDs that are not shared with all the other parties and are not needed for the model training, as they all need to have the same ID set. As privacy must be preserved at all times, parties cannot know which IDs they share and do not share before sending them. Consequently, the PSI server becomes responsible for identifying those common IDs and computing the intersections of only the common IDs. This is achieved by taking each feature of all parties and creating an intersection of all of them, resulting in a list of IDs that all parties share and can be used to train the model.

\begin{algorithm}[t]
\caption{Feature Permutation and PSI Computation}
\label{alg:shapley_cmi_psi}
\begin{algorithmic}[1]
\Require Data parties $X_{all}$, task party $X_t$, task labels $Y_t$
\Ensure PSI values per feature $\textit{psi\_values}$, number of common IDs $|\textit{common\_IDs}|$
\State Combine all features of $X_{all}$ and $X_t$ in a single dictionary
\State \textit{common\_IDs} $\leftarrow$ Identify set of common IDs of all parties
\State Initialize \textit{psi\_values} $\leftarrow\ \emptyset$
\For{$i = 1$ to $N$ permutations}
    \State Generate a random permutation of the feature order
    \State Initialize \textit{computed\_features} $\leftarrow \emptyset$
    \State Initialize set \textit{processed\_IDs} $\leftarrow \emptyset$
    \For{\textbf{each} feature $f$ \textbf{in} permutation}
        \State Get ID groups of $f$ as $X_d$
        \State Get ID groups of \textit{computed\_features}  as $X_D$
        \State Initialize \textit{psi\_results} $\leftarrow \emptyset$
        \For{\textbf{each} $id$ \textbf{in} \textit{common\_IDs}}
            \If{$id \notin \textit{processed\_IDs}$}
                \State $X'_d \: \leftarrow\;$ Group in $X_d$ that contains $id$
                \State $Y'_t \; \, \leftarrow\;$ Group in $Y_t$ that contains $id$
                \State $X'_D \leftarrow\;$ Groups in $X_D$ that contain $id$
                \State Compute intersections:
                \State \quad $A = X'_d \cap X'_D \cap Y'_t \cap \textit{common\_IDs}$
                \State \quad $B = X'_D \cap Y'_t \cap \textit{common\_IDs}$
                \State \quad $C = X'_d \cap X'_D \cap \textit{common\_IDs}$
                \State \quad $D = X'_D \cap \textit{common\_IDs}$
                \State Store $|A|$, $|B|$, $|C|$, $|D|$ in \textit{psi\_results}
                \State Add IDs in $A$ to \textit{processed\_IDs}
            \EndIf
        \EndFor
        \State Append \textit{psi\_results} of $f$ to \textit{psi\_values}
        \State Add $f$ to \textit{computed\_features} 
    \EndFor
\EndFor
\State \Return \textit{psi\_values}, $|\textit{common\_IDs}|$
\end{algorithmic}
\end{algorithm}

After identifying the common IDs, the process of computing all intersections begins by defining the number of random permutations that will be performed, repeating the process as many times as the number of permutations chosen. Features will be added one by one in each permutation until all features are added and the permutation ends. For each feature, it will compute the intersection of all possible value combinations that the IDs have.

As a result, it will retrieve the ID group to which the current ID belongs in the current computing feature, along with the already computed features and the task label, and then determine how many other IDs share the same ID group. For each ID, it will look for the number of other IDs that share the same ID groups under four different conditions: IDs that share all ID groups, IDs that share the ID groups of computed features and task label, IDs that share the ID groups of current feature and computed features, and IDs that share the ID groups of computed features. These four intersections are the ones needed to apply the CMI formula (\eqref{eq:cmi}). The IDs that share all ID groups are the ones that have the same combination of values, so they will not be computed again. This avoids redundant calculations and further improves efficiency, especially in datasets with many repeated patterns. All these intersections will be added to the list of the current permutation of the current feature, repeating the process with the rest of the features in the permutation, and then repeating it all over again throughout all the rest of the permutations.

At the end of all permutations, as a result, a list of features will be obtained that contains a list of permutations, each of which has a list of all the groups of four intersections computed. This data structure allows each client to reconstruct the CMI terms required for the final Shapley-CMI aggregation locally. The number of common IDs they all share will also be obtained, which is necessary to average the final result of the CMI value.

\subsection{Shapley-CMI calculation}
The obtained result needs to be sent back to the parties so that they can compute all the CMI values. Consequently, the PSI server identifies the owner of every feature and sends the PSI values back to them. This mapping ensures that each party receives only the information relevant to its own features, maintaining privacy boundaries. Each party receives them one by one and iterates through each permutation and its combination of values to use the PSI intersection sizes to calculate the CMI value for each one, adding all CMI values together. This local computation is crucial, as the PSI server does not perform any information-theoretic calculations, preserving its role as a non-intrusive coordinator. Afterwards, it averages the total CMI value based on the number of common IDs to obtain the Shapley-CMI value of each feature. Finally, it will add all Shapley-CMI values together to calculate the average at the end, based on the number of permutations, and obtain the actual mean Shapley-CMI value of the entire feature. This final mean captures the expected marginal contribution of the feature across all possible collaboration scenarios. This process will be repeated for every feature the party owns and the PSI server sends back to it, resulting in an estimation of the contribution of every feature of all parties.

\section{Experiment}

\begin{table*}[t]
\centering
\caption{Feature importance comparison across methods}
\label{tab:feature_importance}
\begin{tabular}{l|c|c|c|c|c|c|c|c|c|c|c|c|c}
\textbf{Method} & \textbf{f1} & \textbf{f2} & \textbf{f3} & \textbf{f4} & \textbf{f5} & \textbf{f6} & \textbf{f7} & \textbf{f8} & \textbf{f9} & \textbf{f10} & \textbf{f11} & \textbf{f12} & \textbf{f13} \\
\hline
Shapley-CMI & 
0.0873 & 0.0576 & 0.0339 & 0.0505 & 0.0587 & 0.0745 & 0.1190 & 0.0534 & 0.0532 & 0.1180 & 0.0874 & 0.1134 & 0.0932 \\
Shapley-CMI w/ enc. & 
0.0873 & 0.0576 & 0.0339 & 0.0505 & 0.0587 & 0.0745 & 0.1190 & 0.0534 & 0.0532 & 0.1180 & 0.0874 & 0.1134 & 0.0932 \\
\hline
SHAP & 
0.0989 & 0.0354 & 0.0123 & 0.0046 & 0.0324 & 0.0985 & 0.2925 & 0.0335 & 0.0213 & 0.1055 & 0.0457 & 0.1290 & 0.0904 \\
\end{tabular}
\end{table*}

To validate the effectiveness and security of our implementation, we compared the Shapley-CMI values computed by the original method of \cite{han_data_2021} with our method, which uses encryption with discretized IDs, and the SHAP values obtained from a standard machine learning model. The experiment aims to demonstrate that the privacy-preserving implementation described in this paper yields consistent results with the original Shapley-CMI computation, while also comparing it with a widely used model-dependent data valuation method.

\subsection{Experimental setup}
The experiment was conducted using a vertically partitioned dataset, composed of data shared among three parties, each owning a subset of the features. One of those three parties is the task party, which owns the target label, while the other two are the data parties that provide the additional features for the model training. This setup reflects a typical VFL environment where data features are distributed across different entities. We used a publicly available real-world dataset, Wine \cite{cortez2009modeling}, which contains 178 samples, 13 features, and one label. Those features were distributed evenly across the parties, so that all parties have features to provide. Distributing the features evenly ensures balanced contributions and avoids preferences towards any single party.

For the Shapley-CMI experiments, both the original and encrypted methods were executed under identical settings, including the number of permutations used for feature value estimation and the order of features of each permutation. The reason behind this is that, as the method uses random permutations instead of performing all permutations to reduce computational load, the resulting Shapley-CMI value can vary by some decimals, without indicating that there is something wrong with this approach. Afterwards, all Shapley-CMI values were normalized to know what percentage each feature contributes to the model.

To compute the SHAP values, a model was created by training a simple Random Forest classifier on the whole (centralized) dataset, using the same label as a reference point for comparison. After training the model and ensuring it has a high accuracy, we calculated its Shap values to get the marginal contribution of each feature to the model. Finally, all Shap values were normalized to determine the percentage contribution of each feature to the model and to compare it with the Shapley-CMI values. Normalization facilitates the interpretation and comparison of contributions across features with different scales.

\subsection{Experimental results}
Table \ref{tab:feature_importance} shows the results of Shapley-CMI and SHAP obtained from the experiment. The experiment demonstrates that the feature importance values obtained using the original Shapley-CMI method (based on entropy calculations) and our encrypted version (using PSI) are identical. This demonstrates that the security enhancements do not compromise the accuracy or validity of the feature valuation. In the first case, the CMI values were computed directly from the raw distributions of the features and the label. In contrast, the encrypted version used a secure PSI-based protocol to calculate the required intersection sizes without revealing any sensitive data, using the method described above of discretizing data and creating encrypted ID groups. The fact that both methods give identical results confirms the correctness and reliability of our privacy-preserving implementation.

When comparing the Shapley-CMI values to those obtained using SHAP on a Random Forest model, we observe that both methods identify the most influential features in a similar order. While the values are not the same, mainly because SHAP is model-dependent and Shapley-CMI is not, Shapley-CMI is therefore unable to estimate the exact contribution of each feature. Nevertheless, they still reflect consistent patterns of feature contribution. This shows that even without access to a centralized model, the Shapley-CMI framework (and our secure implementation of it) provides a reasonably accurate estimation of how much each feature contributes to the prediction task.

\section{Discussion}
This method makes new contributions to the process based on \cite{han_data_2021}. A key strength of the proposed system is that it does not need a model to measure the contribution of features, and that it never shares raw data among parties or with the PSI server. All information exchanged consists of encrypted ID sets grouped by discretized feature values, ensuring that no sensitive attributes or labels are exposed at any point in the process. As a result, no party can infer the private data of others. Additionally, it reduces the number of iterations required for the PSI server to compute the intersections of all combinations of values, as it only computes the combinations that the IDs have, rather than computing all of them. This reduces the workload and time needed for the PSI server to compute all the intersections. Finally, the system is highly scalable, supporting any number of parties and features without requiring significant architectural changes. This flexibility makes it applicable to different real-world use cases involving multiple institutions and large datasets.

However, this method also has some limitations. This approach assumes a semi-honest scenario, where all participants are expected to follow the protocol correctly, even if they might try to infer information from the data they receive. This assumption leaves room for potential vulnerabilities if any party or the server behaves maliciously. For example, an adversarial participant could attempt to deviate from the protocol to extract unauthorized information or disrupt the computation. In particular, if the PSI server were malicious and had access to the encryption key, it could carry out brute-force attacks to decrypt IDs back to their original values. Nevertheless, even if they aim to decode the IDs, the most they would obtain would be the original IDs; the real data of their features would remain private, as it is never sent directly to the server. This partial exposure, while limited, still necessitates careful key management and trust assumptions.

Future work aims to continue exploring other privacy-preserving techniques beyond HMAC-SHA256, such as alternative hashing algorithms, homomorphic encryption, or other methods that allow for encrypting data but still computing the intersection. Furthermore, defences against a malicious PSI server should be considered, including distributed or verifiable PSI protocols that reduce the trust placed in a single central server. These improvements would make the system more resilient in adversarial environments and increase its applicability to stricter privacy contexts.

\section{Conclusions}
This paper presents a secure and scalable implementation of Shapley-CMI for VFL, based on an existing information-theoretic method for data valuation that does not rely on any machine learning model. Our contributions focus on preserving privacy through encryption with HMAC-SHA256 and achieving efficiency and scalability by discretizing features and structuring the PSI protocol accordingly. These improvements enable the system to compute the permutations needed to measure Shapley values, which estimate the contribution of features, without directly sharing raw data. The experimental results validate the correctness and robustness of the proposed method by demonstrating that the original Shapley-CMI results and those with encryption are identical. Additionally, it provides an accurate estimation of the SHAP values, which measure the actual contribution each feature makes to the model. The work demonstrates the feasibility and importance of privacy-preserving mechanisms in collaborative learning environments, especially when sharing the raw data is not an option.

\bibliographystyle{ieeetr}
\bibliography{References_IEEE}

\end{document}